\documentclass[twocolumn]{aastex631}
\usepackage{placeins}
\usepackage{amsmath} 
\usepackage{multirow}
\usepackage{subfigure}
\newcommand{\new}[1]{\textcolor{black}{{\bf{#1}}}}

\begin{document}

\title{Self-similar solutions of oscillatory reconnection: parameter study of magnetic field strength and background temperature}

\author[0000-0002-5082-1398]{Luiz A. C. A. Schiavo}
\affiliation{Northumbria University, Newcastle upon Tyne, NE1 8ST, UK}

\author[0000-0002-5915-697X]{Gert J. J. Botha}
\affiliation{Northumbria University, Newcastle upon Tyne, NE1 8ST, UK}

\author[0000-0002-7863-624X]{James A. McLaughlin}
\affiliation{Northumbria University, Newcastle upon Tyne, NE1 8ST, UK}



\begin{abstract}
{
Oscillatory reconnection is a specific type of time-dependent reconnection which involves periodic changes in the magnetic topology of a null point. The mechanism has been reported for a variety of magnetic field strengths and configurations, background temperatures and densities. All these studies report an oscillation in the current density at the null point, but  also report a variety of periods, amplitudes and overall behaviors.  We conduct a parametric study for equilibrium magnetic field strength and initial background temperature, solving 2D resistive MHD equations around a magnetic X-point. We introduce a parameter space for the ratio of internal-to-magnetic energy and find self-similar solutions for simulations where this ratio is below 0.1 (which represents a magnetically-dominated environment or, equivalently, a low-beta plasma). Self-similarity can be seen in oscillations in the current density at the null (including amplitude and period), Ohmic heating and the temperature generated via reconnection jets. The parameter space of energy ratios also allows us to contextualize previous studies of the oscillatory reconnection mechanism  and bring those different studies together into a single unified understanding.
}
\end{abstract}



\section{Introduction} \label{sec:intro}

Magnetic reconnection is a fundamental plasma process allowing stored magnetic energy to be released into thermal and kinetic energy, as well as accelerate  particles and allow a change in magnetic connectivity (\citealt{2022LRSP...19....1P}; \citealt{BROWNING2024100049}). Reconnection is understood to be at the heart of several fundamental processes, including coronal mass ejections (e.g. \citealt{2012LRSP....9....3W}) and solar flares (e.g. \citealt{2017LRSP...14....2B}). For example, observations of chromospheric anemone jets have provided evidence of reconnection events occurring  at smaller spatial scales in the chromosphere, suggesting a potential link between the heating of the solar chromosphere and corona and small-scale reconnection \citep{Shibata2007}.
%


Oscillatory reconnection is a specific type of time-dependent reconnection which involves periodic changes in the magnetic connectivity and topology of the field. The concept of oscillatory reconnection was first identified by \cite{Craig1991} during their investigation of the relaxation of a two-dimensional (2D) X-point configuration magnetic field. One distinguishing feature of oscillatory reconnection is its intrinsic periodicity, which arises naturally from the relaxation process itself rather than being externally imposed (i.e. the generation of periodic outputs even from aperiodic drivers \citealt{McLaughlin2012a}).

%

\cite{McLaughlin2009} investigated a 2D X-point configuration in a cold plasma simulating the resistive magnetohydrodynamic (MHD) equations for a ideal fully-ionized plasma. In this study, oscillatory reconnection was initiated by perturbing the magnetic X-point using an external fast magnetoacoustic pulse. The research identified several key properties of this mechanism, including periodic changes in the orientation of the resulting current sheet corresponding to alterations in magnetic connectivity. Additionally, the formation of both fast and slow oblique magnetic shocks was observed as part of the oscillatory reconnection process. Studies have also highlighted the role of oscillatory reconnection in generating quasi-periodic waves and flows in the solar atmosphere, providing a physical explanation for high-speed, quasi-periodic, transverse outflows and jets \citep{McLaughlin2012b}.

Investigations into the three-dimensional (3D) nature of oscillatory  reconnection have revealed that reconnection at fully 3D nulls can occur naturally in a time-dependent and periodic fashion \citep{Thurgood2017}. The periodicity of oscillatory reconnection has been found to be independent of the initial pulse in 2D X-point simulations \citep{Karampelas2022b}. More recently, \cite{Talbot2024} investigated the impact of resistivity on oscillatory reconnection, and  discovered that the reconnection period is independent of background resistivity. Additionally, simulations of emerging flux in a coronal hole have demonstrated the initiation of oscillatory reconnection in a self-consistent manner, with signatures comparable to observations of select flux emergence events and solar and stellar flares \citep{Murray2008}. Moreover, oscillatory reconnection has been proposed as a possible mechanism underlying various periodic phenomena in the solar atmosphere, including quasi-periodic pulsations (QPPs, e.g. \citealt{McLaughlin2018}; \citealt{Zimovets2021}).

Oscillatory reconnection phenomena have been investigated across various plasma configurations. Among these studies, \cite{Prokopyszyn2019} delved into the dynamics of a null point perturbed by a continuous driver perpendicular to a 2D plane that contains an X-point using 2.5D simulations, revealing phase-mixing due to the magnetic field inhomogeneities. \cite{Santamaria2018} examined oscillatory reconnection within a 2D arcade configuration, considering a stratified atmosphere, finding that the null point behaves as a resonant cavity generating waves at certain frequencies that depend upon the equilibrium parameters. \cite{Tarr2017} studied a null point in an arcade configuration, modeling a stratified atmosphere from the photosphere to the lower corona. In their simulation, they added a wave packet driver in the photosphere. They analyzed the energy conversion of this incident wave packet at the null point. They reported that 70\% of the energy incident on a null point is converted to slow magnetoacoustic waves, 7\% into fast magnetoacoustic waves, and 23\% remains at the null until dissipated. \cite{Stewart2022}  found that, during flux rope coalescence, oscillatory reconnection can occur intrinsically without an external oscillatory driver, resulting in both a periodic signal and the generation of radially-propagating nonlinear waves. In these investigations \citep{Prokopyszyn2019,Santamaria2018,Stewart2022,Tarr2017}, the plasma was either analyzed under fixed atmospheric conditions or within a single magnetic field configuration.

%

\cite{Karampelas2022a} investigated the periodicity and decay rate in the oscillatory reconnection pattern observed at the current sheet at a null point. In their study, the effect of temperature was evaluated from 0 K up to 1 MK for a 1.44 G magnetic field. Their findings show that the oscillatory reconnection signal is only affected by temperatures above 10,000 K. This study's parametric exploration of temperature and magnetic field variations was conducted within a limited parameter space. \cite{Karampelas2023} performed a parametric study for evaluating the impact of solar atmospheric conditions and magnetic field on the oscillatory reconnection period. They evaluated a temperature range from 3 MK up to 10 MK for a magnetic field range from 10 G to 30 G and proposed an empirical formula to describe the oscillatory reconnection period.

Thus, oscillatory reconnection has been studied in a variety of magnetic topologies, albeit all containing a null point, and for a variety of coronal conditions, including different initial temperature profiles and varying equilibrium magnetic field strengths. All these studies report an oscillating signal in the current density at the null point; a tell-tale sign of oscillatory reconnection. However, these studies also report a variety of periods, amplitudes and behaviors for such a signal, and it is currently unclear how to bring these different studies together into a  single unified understanding.  This paper aims to do just that:  to perform a parametric study involving variations in magnetic field strength from $B =$ 5 G to 100 G and plasma temperature between 0 K to 10 MK, with the aim of exploring a wider parameter space than previous explored, in order to consolidate these different results under a single explanation.

This paper has the following structure: the numerical model, initial conditions and boundary conditions are detailed in $\S$\ref{sec:numerical-model}; the results are presented in $\S$\ref{sec:results}, including the dependence on the equilibrium magnetic field ($\S$\ref{sec:b-influence}), the 
influence of the initial background temperature  ($\S$\ref{sec:TEMP-influence}) and the unification of these results into an energy map ($\S$\ref{sec:energy-map}). The conclusions are presented in $\S$\ref{sec:conclusions}.




\begin{table}[t]
\centering
\caption{Initial conditions employed in the parametric study, with values given at 1 Mm from the null point.}
 \label{tab:cases}
\begin{tabular}{lrrrrrrr}\hline \hline
 \multirow{2}{*}{Case} & \multirow{1}{*}{$B_0$} & \multirow{1}{*}{$T_0$} & \multirow{1}{*}{$p_0$} & \multirow{2}{*}{$\displaystyle\frac{V_{s0}}{V_{a0}}$} & \multirow{2}{*}{$\beta_0$} & \multirow{2}{*}{$\displaystyle\frac{E_{i0}}{E_{B0}}$}  \\
                       & \multirow{1}{*}{(G)}   & \multirow{1}{*}{(MK)}  & \multirow{1}{*}{(Pa)}  & \multirow{1}{*}{} & \multirow{1}{*}{} & \multirow{1}{*}{} \\ \hline 
        A1 & 5 & 10 & 2.76E-1 & 1.52 & 6.24 & 4.16 \\
        A2/D1  & 10 & 10 & 2.76E-1 & 0.76 & 1.56 & 1.04 \\ 
        A3 & 20 & 10 & 2.76E-1 & 0.38 & 0.39 & 0.26 \\ 
        A4/C1 & 50 & 10 & 2.76E-1 & 0.15 & 0.062 & 0.042 \\ 
        A5/B1 & 100 & 10 & 2.76E-1 & 0.076 & 0.016 & 0.010 \\ \hline
        B2 & 100 & 1 & 2.76E-2 & 0.024 & 1.6E-3 & 1.0E-3 \\ 
        B3 & 100 & 0.1 & 2.76E-3 & 7.6E-3 & 1.6E-4 & 1.0E-4 \\ 
        B4 & 100 & 0.001 & 2.76E-5 & 7.6E-4 & 1.6E-6 & 1.0E-6 \\ 
        B5 & 100 & 0 & 0 & 0 & 0 & 0 \\ \hline
        C2 & 50 & 1 & 2.76E-2 & 0.048 & 6.2E-3 & 4.2E-3 \\ 
        C3 & 50 & 0.1 & 2.76E-3 & 0.015 & 6.2E-4 & 4.2E-4 \\ 
        C4 & 50 & 0.001 & 2.76E-5 & 1.5E-3 & 6.2E-6 & 4.2E-6 \\ 
        C5 & 50 & 0 & 0 & 0 & 0 & 0 \\ \hline
        D2 & 10 & 1 & 2.76E-2 & 0.24 & 0.16 & 0.10 \\ 
        D3 & 10 & 0.1 & 2.76E-3 & 0.076 & 0.016 & 0.010 \\ 
        D4 & 10 & 0.001 & 2.76E-5 & 7.6E-3 & 1.6E-4 & 1.0E-4 \\ 
        D5 & 10 & 0 & 0 & 0 & 0 & 0 \\ \hline
\end{tabular}
 \end{table}

\section{Numerical model} \label{sec:numerical-model}
\subsection{Governing equations}
In our investigation, we solve the 2D resistive MHD equations through the utilization of the LARE2D code \citep{Arber2001}. The equations are solved in Lagrangian form, employing a Lagrangian-Eulerian remap procedure and can be expressed in dimensionless form as follows:

\begin{eqnarray*}
\frac{D\rho}{D t} &=& - \rho \nabla \cdot \mathbf{v} , \\
\frac{D\mathbf{v}}{D t} &=& \frac{1}{\rho}(\nabla \times \mathbf{B} ) \times \mathbf{B} - \frac{1}{\rho}\nabla p  , \\
\frac{D\mathbf{B}}{D t} &=& (\mathbf{B}\cdot\nabla) \mathbf{v} - \mathbf{B}(\nabla \cdot \mathbf{v}) - \nabla\times (\eta\nabla\times \mathbf{B}) , \\
\frac{D\epsilon}{D t} &=& - \frac{p}{\rho} \nabla \cdot \mathbf{v} + \frac{\eta}{\rho}|\mathbf{j}|^2 , \\
p &=& \rho \epsilon(\gamma -1) .
\label{eq:mhd}
\end{eqnarray*}
Here, $\mathbf{v}$ denotes the velocity vector, $\mathbf{B}$ represents the magnetic field, $\mathbf{j}$ is the current density, $\rho$ signifies plasma density, $p$ corresponds to plasma thermal pressure, $\epsilon$ represents specific internal energy, $\eta$ characterizes the resistivity, and $\gamma$ is the ratio of specific heats, set to 5/3 for a hydrogen plasma. To accurately accommodate steep gradients like shocks and address numerical instabilities, LARE2D utilizes a numerical viscosity \citep{Caramana1998,Arber2001}.

The model assumes full ionization of the plasma and non-dimensionalizes the governing equations with respect to length-scale $L_0$, magnetic field $B_0$, and density $\rho_0$. These constants define non-dimensionalization for velocity $v_0 = B_0/\sqrt{\left.\mu_0 \rho_0\right.}$, thermal pressure $p_*=B_0^2/\mu_0$, time $t_0=L_0/v_0$, current density $j_0=B_0/\mu_0 L_0$, specific internal energy $\epsilon_0=v_0^2$, temperature $T^*=\epsilon_0\overline{m}/k_B$ and resistivity $\eta_0= \mu_0L_0v_0$, where $\mu_0$ is the vacuum magnetic permeability, $k_B$ is the Boltzmann constant and $\overline{m}$ the average mass of ions. Simulation results can be scaled with appropriate reference scales, with typical values for the solar corona being $L_0 =$ 1 Mm and $\rho_0 = 1.67 \times 10^{-12}$ kg/m$^{3}$. We set the resistivity as $\eta=10^{-4}\eta_0$. Our investigation explores a variety of $B_0$ values and initial background temperatures, and the physical values of these are presented in Table \ref{tab:cases}. There is no physical viscosity in our system and the numerical dissipation is negligible. It is important to notice that some normalization scales depend on $B_0$, as shown in Table \ref{tab:normalisation}. 

We will conduct two types of parametric studies: simulation set A, where the temperature and thermal pressure are initially constant while varying the magnetic field intensity, and simulation sets B, C and D, where we choose and fix an equilibrium magnetic field strength $B_0$, while exploring increasing the initial plasma temperature from a cold state ($T=0$) up to 10 MK, for cases at $B_0=$ 10 G, 50 G and 100 G. Details of each simulation set is shown in Table \ref{tab:cases}.

In our analysis, we consider a fully-ionized, pure hydrogen plasma, wherein the average mass of ions can be approximated to the proton mass, $m_p$. However, for conditions typical of the solar corona, where the plasma composition includes various elements, one can account for an average ion mass by setting $\overline{m}=1.2m_p$. Due to the non-dimensionalization in our system, it is straightforward to consider either a pure hydrogen plasma ($\overline{m}=m_p$, where these are the results presented in this paper) or the temperature derived from our simulations can be divided by 1.2 to obtain the corresponding temperature for an average ion mass of $\overline{m}=1.2m_p$ (where this adjustment would then account for heavier ions in the plasma composition).



\subsection{Equilibrium magnetic field and initial condition in velocity}\label{sec:2.2}
For the equilibrium magnetic field, we consider a 2D X-point defined by:
\begin{eqnarray}
\mathbf{B}=\frac{B_0}{L_0}(y,x,0) .\label{equation=X-point}
\end{eqnarray}

\begin{table}[t]
\centering
\caption{Non-dimensionalization scales showing the influence of magnetic field strength $B_0$ on time $t_0$, temperature $T^*$ and current densities $j_0$.}
 \label{tab:normalisation}
\begin{tabular}{cllll}\hline \hline
Case & $B_0$ (G) & $t_0$ (s) & $T^*$ (MK) & $j_0$ (A/m2)     \\ \hline 
A5, B1-B5 & 100    & 0.145  & 5772.8  & 7.957E-3  \\
A4, C1-C5 & 50     & 0.290  & 1443.2  & 3.978E-3  \\
A3 & 20     & 0.724  & 230.9   & 1.591E-3  \\
A2, D1-D5 & 10     & 1.449  & 57.7    & 0.795E-3 \\
A1 & 5      & 2.897  & 14.4    & 0.397E-3 \\ \hline
\end{tabular}
\end{table}

\noindent We take the equilibrium density, $\rho_0$, and initial background temperature, $T_0$, to be uniform, and the magnetic Reynolds number was set as $R_m = 10^4$. The initial temperature profile and equilibrium magnetic field strength are detailed in Table \ref{tab:cases}. Table \ref{tab:cases} also shows information on the ratio between internal energy $E_i=\rho \epsilon$, and magnetic energy $E_B = |\mathbf{B}|^2/2\mu_0$, per unit of volume. $V_s=\sqrt{\gamma p/\rho}$ is the speed of sound and $V_a=|\mathbf{B}|/\sqrt{\mu_0 \rho}$ is the Alfv\'{e}n speed. 
Note that the equilibrium magnetic field given in Equation (\ref{equation=X-point}) is highly inhomogenous and is scale-free. Thus, our choice of $B_0$ is only the initial value of magnetic field strength at $t=0$ and $r=L_0$, where $r=\sqrt{x^2+y^2}$, i.e., $|\mathbf{B}(r=L_0,t=0)|=B_0$. Similarly,  $V_a(r=L_0,t=0)=V_{a0}$, $E_i(r=L_0,t=0)=E_{i0}$ and  $E_B(r=L_0,t=0)=E_{B0}$ denote the initial Alfv\'{e}n speed, internal energy per unit of volume, and magnetic energy per unit of volume, respectively, at $r=L_0$ and $t=0$. In the same way, $T(t=0)=T_0$, $p(t=0)=p_0$ and $V_s(r=L_0,t=0)=V_{s0}$ denote the initial background temperature, pressure and sound speed, respectively, which we take to be constant when $t=0$ (and thus there is no need to specify $r=L_0$). All these parameter choices are detailed in Table \ref{tab:cases}.

The plasma $\beta$ denotes the ratio of thermal pressure ($p$) to magnetic pressure ($p_{\rm{magnetic}}=|\mathbf{B}|^2/2\mu_0$) and is given by $\beta= 2\mu_0 p / |\mathbf{B}|^2 $. Initially and at $r=L_0$, we define $\beta_0$ as:
\begin{equation}
    \beta_0 = \beta|_{r=L_0, t=0} = 
    \left.\frac{2\mu_0 p}{|\mathbf{B}|^2} \right|_{r=L_0, t=0} =
    \frac{2\mu_0 p_0}{B_0^2} \ .
    \label{eq:beta0}
\end{equation}



\begin{figure}[t]
 \centering
	\includegraphics[width=0.99\columnwidth]{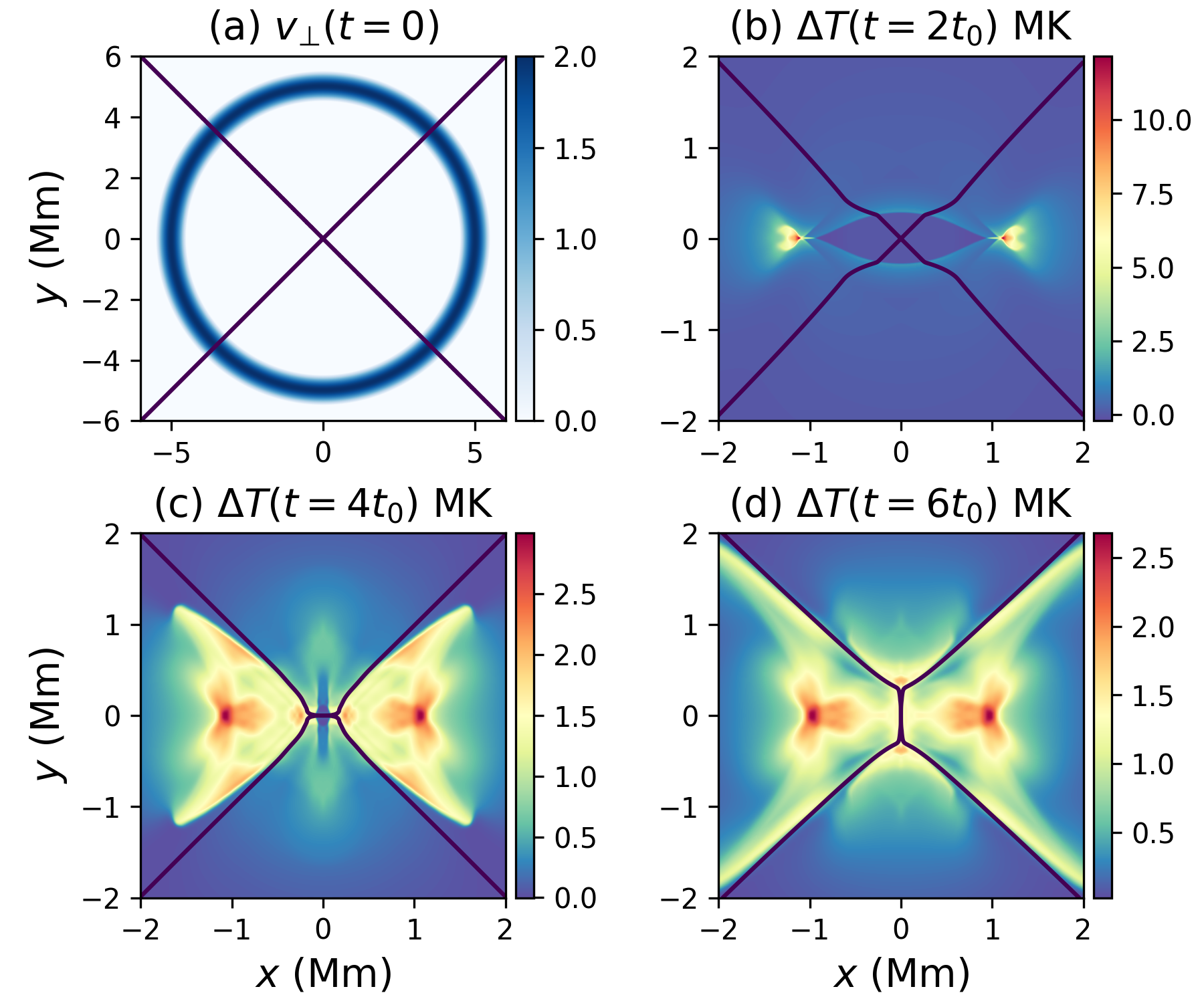}
	\caption{Evolution of the plasma flow for case D2, where $B_0 = $ 10 G and $T_0=$ 1 MK. Panel (a) shows contour of $v_\bot$ and separatrices, panels (b) to (d) present contours of $\Delta T=T-T_0$ and black lines represent the (evolving) separatrices.}
	\label{fig:flow-evo}
\end{figure}

The initial velocity field is computed as it was in  \citet{McLaughlin2009}, where it is imposed based in two variables  $v_\bot = \mathbf{v} \times \mathbf{B}   \cdot \hat{\mathbf{z}}$ and $v_\parallel=\mathbf{v}\cdot\mathbf{B}$ that is related to propagation perpendicular and parallel to the magnetic field lines. The initial velocity pulse is given by:
\begin{eqnarray}
	v_\bot(x,y) &=& 2C\mbox{sin}[\pi(r-4.5) ] \hspace{0.3cm} 4.5 \le r\le 5.5 , \label{equation=v_perp}\\
	v_\parallel(x,y) &=& 0 \label{equation=v_parallel} ,
\end{eqnarray}
where 2$C$ is our initial amplitude. The expression describe a circular, sinusoidal pulse shown in Figure \ref{fig:flow-evo}a. When the simulation begins, this initial pulse will naturally split into two waves, each of amplitude $C$, traveling in different directions: a radially-outgoing wave and a radially-incoming wave. The incoming wave, i.e. the wave traveling towards the null point, is the wave we are primarily interested in since it is the wave that is responsible for triggering the oscillatory reconnection. The Cartesian velocity field, equivalent to Equations (\ref{equation=v_perp}) and (\ref{equation=v_parallel}),  can be obtained via: 
\begin{eqnarray*}
v_x = \frac{v_\parallel B_x +v_\bot B_y}{|\mathbf{B}|^2} \; \;{\rm{and}} \; \; v_y = \frac{v_\parallel B_y -v_\bot B_x}{|\mathbf{B}|^2}. 
\end{eqnarray*}


\subsection{Boundary conditions and domain setup}
We adopt a Neumann boundary condition imposing zero gradient at the boundaries for velocities, magnetic field and thermodynamic variables. We also employ a stretched grid characterized by finer resolution closer to the null point, i.e. the region of primary interest, and coarser resolution in the outer regions. The grid is equally spaced and highly refined around the null point and initial pulse at -5 $\le x,y\le$ 5, and it employs a stretching at $ x,y>$ 5. We adopt a hyperbolic tangent stretching function that smoothly changes the growth rate of grid spacing from 0 to 7\% for $ x,y>$ 5. The mesh stretching in the outer regions also creates some numerical dissipation, which is useful in terms of dissipating away the outgoing waves and thus reducing the impact of reflected waves which could then go on to further perturb our null point, which is undesirable. We also employed a damping region at $r>6$, such that this damping region removes kinetic energy from the outgoing waves in the outer region, again so it does not reflect back and perturb the null. The details of this kinetic-energy damping condition are described in \cite{Talbot2024}. Our total grid has 1700 $\times$ 1700 points, and the total domain box extends to -93 $\le x,y\le$ 93.


\begin{figure}[t]
 \centering
	\includegraphics[width=0.99\columnwidth]{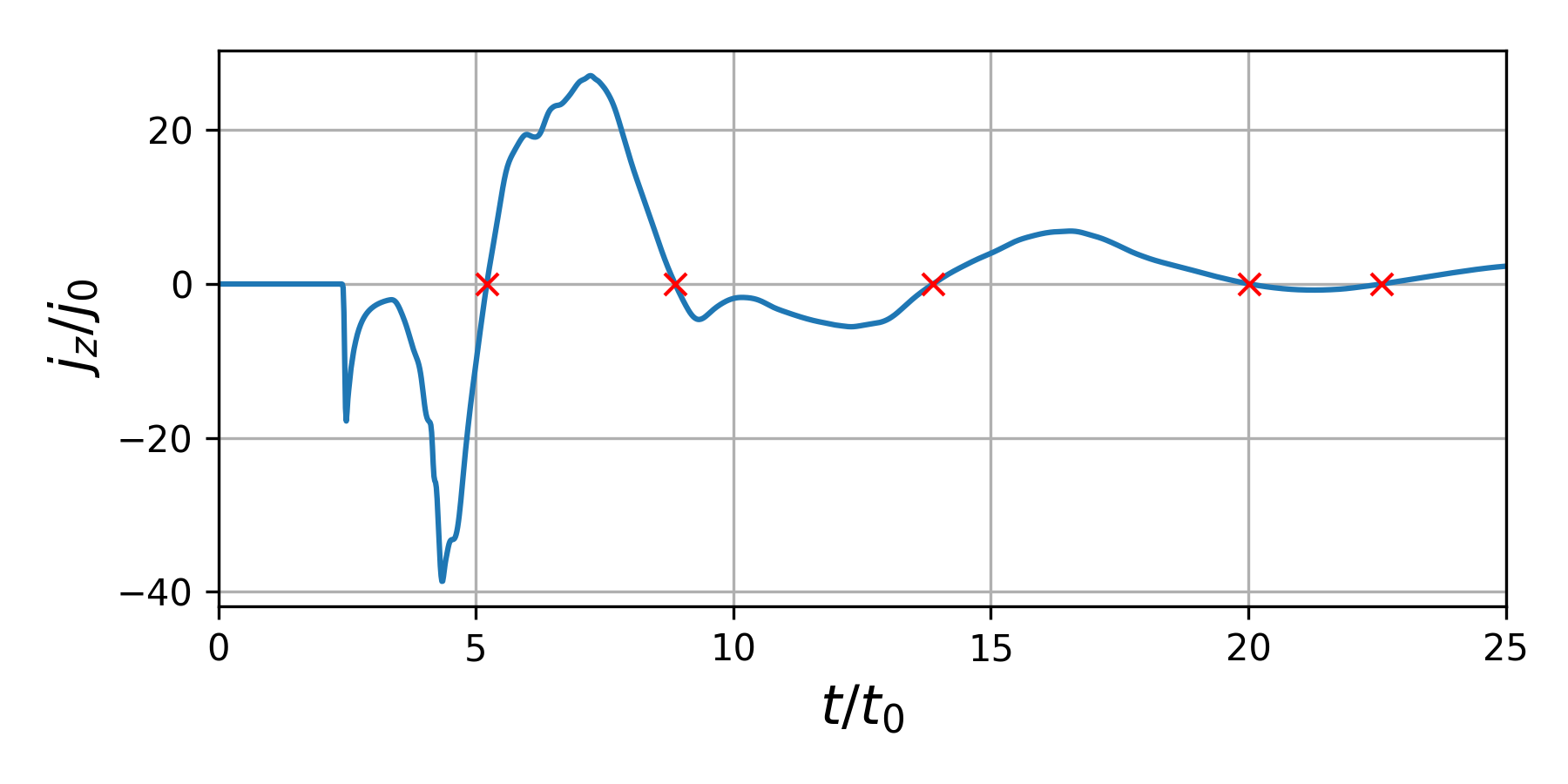}
	\caption{\new{Time evolution of the current density at the null point for case D2, $j_z(0,0,t)/j_0$, with the $\times$ symbol denotes the roots of the function.}}
	\label{fig:flow-evo2}
\end{figure}

\begin{figure}[b]
\centering
\includegraphics[width=0.99\columnwidth]{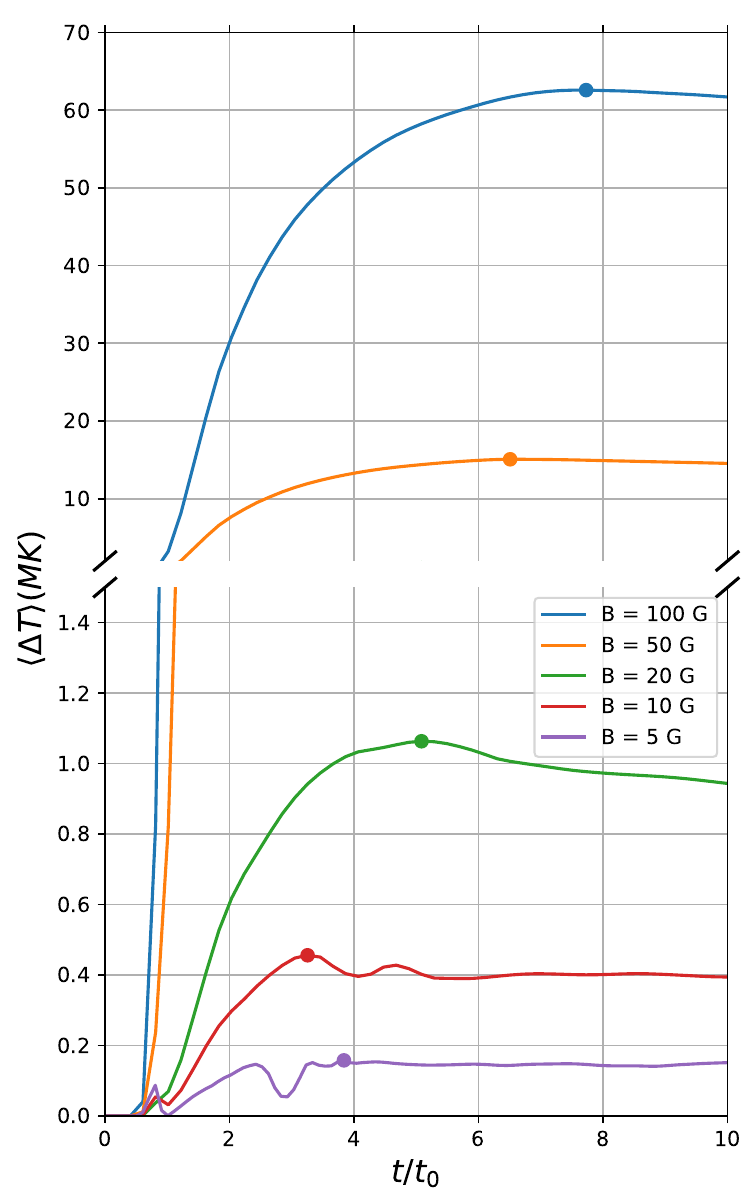}
\caption{Integrated temperature perturbation $\langle \Delta T \rangle$ as function of non-dimensionalized time for simulation sets A1 to A5 in Table \ref{tab:cases}. The circles indicate the maximum of each time series.}
\label{fig:heating-B}
\end{figure}

\section{Results} \label{sec:results}



\subsection{Overall behavior and temperature evolution}

The numerical set-up, choice of equilibrium magnetic field and initial velocity perturbation 
closely follow the work of previous authors (such as \citealt{McLaughlin2009, Karampelas2022a, Karampelas2023}) and readers are referred to these works  for a detailed explanation of the system evolution. Instead, this section will focus on previously unexplored details, such as the analysis of the evolution of temperature perturbation.

Figure \ref{fig:flow-evo} depicts the evolution for simulation D2 in Table \ref{tab:cases}, where $B_0 = $ 10 G and $T_0=$ 1 MK (There is nothing special about the choice of case D2; it just represents a typical simulation across all our cases). In the Figure \ref{fig:flow-evo}a, the initial condition is shown by a circular pulse represented by $v_\bot$ contours, which generates two fast magnetoacoustic waves,
one propagating inward and the other outward. The inward wave perturbs the null point and initiates oscillatory reconnection. Note that, in order to focus attention towards the area of primary interest (the null) the numerical domain presented here is a subset of the full domain.

Subsequent Figures \ref{fig:flow-evo}b -\ref{fig:flow-evo}d illustrate temperature perturbation contours $\Delta T$, with black lines representing the magnetic field lines separatrices. We define $\Delta T$ as the temperature difference between the evolving temperature field and the initial background temperature (which is initially  uniform):
\begin{equation}
\label{eq:dt}
\Delta T(x,y,t) = T(x,y,t) - T(x,y,0) = T(x,y,t) - T_0\ . 
\end{equation}

\begin{figure*}[t]
\centering
\begin{subfigure}
{\includegraphics[trim = 12 10 10 10, clip, width=0.99\columnwidth]{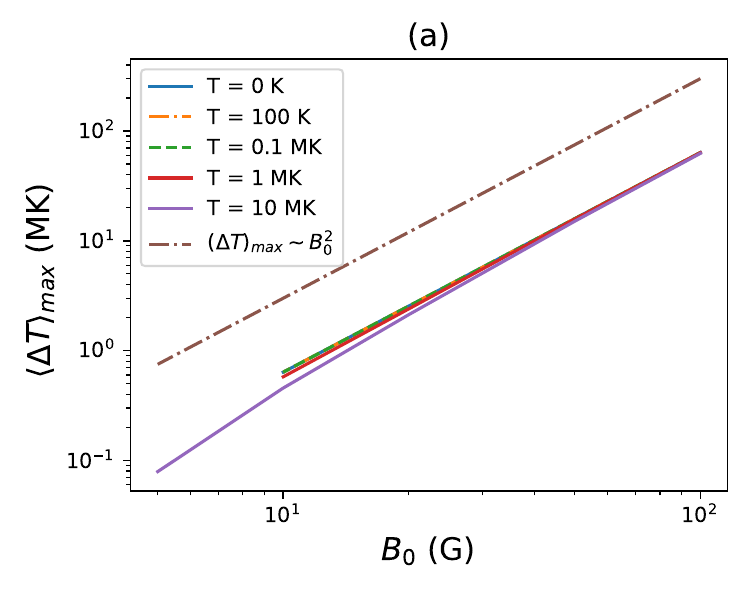}}
\end{subfigure}
\begin{subfigure}
{\includegraphics[trim = 12 10 10 10, clip,width=0.99\columnwidth]{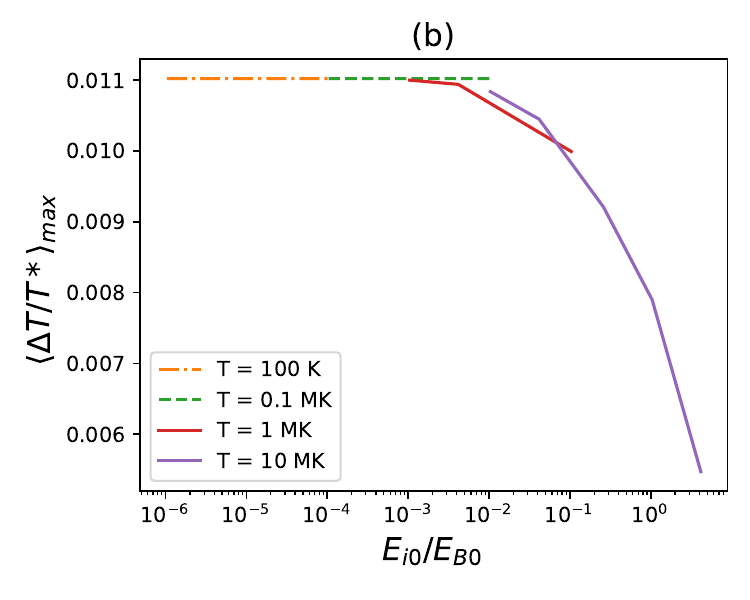}}
\end{subfigure}
\caption{(a) maximum integrated temperature $\langle \Delta T \rangle_{\text{max}}$ as function of equilibrium  magnetic field strength $B_0$. The brown dot-dash line denotes a fitted quadratic dependence on the magnetic field strength following $\langle \Delta T \rangle_{\text{max}} \sim B_0^2$, where the slope is offset artificially for ease of comparison. (b) maximum integrated non-dimensional temperature $\langle \Delta T \rangle_{\text{max}}$ as function of internal-to-magnetic energy ratio. The purple line represents the maximum obtained for simulation set A seen in Fig.\ \ref{fig:heating-B}. The red line represents the maximum from simulations B2, C2 and D2, the green line corresponds to simulations B3, C3 and D3, the orange line cases B4, C4 and D4 and the blue B5, C5 and D5. 
}
\label{fig:heating-loglog}
\end{figure*}

In Figure \ref{fig:flow-evo}b, at $t=2t_0$, the fast magnetoacoustic wave approaches the null point and develops into a shock wave. This initial shock wave elevates plasma temperature and gives rise to two jet streams along the $x-$axis, emanating from the null point. The plasma attains a highly-localized temperature of 12 MK at the jets.

In Figure \ref{fig:flow-evo}c, $\Delta T$ is presented at $t=4t_0$, after the shock wave reaches the null point. The X-point is highly deformed from its equilibrium profile (due to the passage of the fast oblique magnetic shocks) and the magnetic field lines assume a new configuration featuring a horizontal current sheet aligned with the $x-$axis, marking the first cycle of oscillatory reconnection. Additionally, the temperature peak of 12 MK decreases and spreads near the jets, heating the plasma locally to 2.5 MK.

Figure \ref{fig:flow-evo}d, at $t=6t_0$, presents the plasma after the first horizontal current sheet has formed, force imbalance has then peeled the horizontal current sheet apart, overshoot the original magnetic configuration, and so a new current sheet has formed (the first vertical current sheet), parallel to the $y-$axis.  This process characterizes an oscillation cycle, where the reconnection and reorientation will repeat at the points where the roots of the oscillations of $j_z(0,0,t)$ are displayed in Figure \ref{fig:flow-evo2}. In Figure \ref{fig:flow-evo}d, we observe local heating at the ends of the vertical current sheet, albeit the localized heating is still strongest close to $x \approx \pm 1, y=0$. In other words, the strongest localized heating comes from that first horizontal current sheet.




\subsection{Sensitivity to choice of equilibrium magnetic field strength}\label{sec:b-influence}

In this section, we shall analyze the influence of magnetic field strength on heating for a given initial background temperature. To achieve this, we utilize the integrated temperature perturbation $\langle \Delta T \rangle$, to measure the heating effect near the null point. The symbol $\langle \, \, \rangle$ denotes a spatial average over an area $S$, defined as:

\begin{equation}
	\langle f \rangle = \frac{1}{S}\iint_{S} f(x,y,t) dS, 
\end{equation}
where $f$ is some arbitrary function, and $S$ the integration surface we defined as $-2< x,y<2$ Mm. 
The integration area was selected to enclose the magnetic field lines, jet stream and the heated region, as depicted in Figure \ref{fig:flow-evo}.
The integrated temperature perturbation serves as a metric for assessing the heating effect in the vicinity of the null point. The circles in Figure \ref{fig:heating-B} represent the maximum heating observed in the time series.



Figure \ref{fig:heating-B} illustrates the values of $\langle \Delta T \rangle$ for simulation sets A1-A5, as outlined in Table \ref{tab:cases}, wherein the plasma is maintained at 10 MK and we vary the equilibrium magnetic field strength, $B_0$. As mentioned in Section \ref{sec:2.2}, Equation (\ref{equation=X-point}) is scale-free and so our choice of $B_0$ is  the value magnetic field strength at $r=L_0$ (and then $\mathbf{B}$ varies throughout the domain).

In Figure \ref{fig:heating-B}, initially a sharp increase in plasma temperature is observed, coinciding with the passage of the shock wave propagating towards the null point. This initial heating phase concludes approximately before $t =2t_0$. Subsequently, oscillatory reconnection processes sustain the heating by converting magnetic energy into internal energy.

In cases A1, A2 and A3, corresponding to $B_0 = 5$ G, $B_0 = 10$ G and $20$ G respectively, a spike in heating is observed around $t =$1.5$t_0$, attributed to the initial shock wave. Following this, the plasma undergoes cooling before the onset of oscillatory reconnection, heating up to 0.19 MK,  0.48 MK and 1.06 MK, respectively. These cases are characterized by a high $\beta_0$ regime, as indicated by the initial internal-to-magnetic energy ratio of 4.16 for case A1, 1.04 for case A2, and 0.26 for case A3 (at a distance of 1 Mm from the null point). Consequently, the magnetic field strength in cases A1 and A2 is insufficient to heat the plasma to temperatures exceeding 1.0 MK and never above 1.1 MK for case A3.

Conversely, in cases A4 and A5, corresponding to $B_0 = 50$ G and 100 G respectively, the plasma is in a low $\beta_0$ regime, resulting in more pronounced heating. Maximum temperatures reach up to 15 MK and 63 MK, respectively, indicating a more significant heating effect due to the stronger magnetic fields.


Figure \ref{fig:heating-loglog} illustrates the maximum heating obtained from all the
different simulation sets in Table \ref{tab:cases}, namely A1-to-A5, B1-to-B5, C1-to-C5 and D1-to-D5 (note that B1=A5, C1=A4 and D1=A2) cases. Observing Figure \ref{fig:heating-loglog}a, it becomes apparent that the heating is directly proportional to the choice of equilibrium magnetic field strength at 1 Mm from the null point. 
The maximum heating follows a power law and exhibits a quadratic dependence on the magnetic field strength following a $\langle \Delta T \rangle_{\text{max}} \sim B_0^2$ line slope.

\begin{table}[b]
\caption{Fitting coefficients and standard deviation for each curve from Figure \ref{fig:heating-loglog}a, coefficients $a$ and $b$ are presented in Equation (\ref{eq:fitting}).}
\label{tab:fitting}
\centering
\begin{tabular}{ccccc}\hline \hline
		$T$ (MK) & $a $  & $b$       & $\sigma_a$ & $\sigma_b$  \\ \hline
		10     & 4.77E-3 & 2.06 & 2.25E-4  & 1.03E-2 \\
		1      & 6.11E-3 & 2.01 & 1.21E-4  & 4.34E-3 \\
		0.1    & 6.36E-3 & 2.00 & 1.86E-10 & 6.41E-9 \\
		0.001  & 6.36E-3 & 2.00 & 1.86E-10 & 6.41E-9 \\
		0      & 6.36E-3 & 2.00 & 1.86E-10 & 6.41E-9 \\ \hline
\end{tabular}
\end{table}

We performed a fitting on the curves shown in Figure \ref{fig:heating-loglog}a and we were able to obtain an expression to quantify the heating and the average temperature as: 
\begin{equation}    
	\langle \Delta T \rangle_{\text{max}} =  a B_0^{b} , 
 \label{eq:fitting}
\end{equation}
where in this expression $B_0$ is the initial magnetic field in Gauss at 1 Mm from the null point and the initial output temperatures are given in MK. The constants $a$ and $b$ are presented in Table \ref{tab:fitting} together with respective standard deviations, $\sigma_a$ and $\sigma_b$.
Under the assumption that the covariance $\sigma_{ab}$ between the coefficients $a$ and $b$ is negligible, we can derive the following expression to describe the standard deviation for the fitted expression using uncertainty propagation:
\begin{eqnarray*}
	\sigma_{\Delta T} &\approx& a B_0^{b} \sqrt{\left( \frac{\sigma_a}{a}\right) ^2 + 
		B_0^{2b} (\mbox{ln}(B_0)\sigma_b)^4 	 }\;\; .
\end{eqnarray*}

\noindent Table \ref{tab:fitting} shows that there is no variation in the index $b$, or amplitude $a$, for cases with temperatures lower than 1 MK. For cases at 1 MK, there is a negligible variation in the coefficients, and for the hotter case at 10 MK, we can observe a slight variation. The curve at 10 MK comprises the simulation set A1 to A5 (representing the peaks observed in Figure \ref{fig:heating-B}), which has a significant variation in $\beta_0$, ranging from 0.016 to 6.24. The other curves have low $\beta_0$ values, as shown in Table \ref{tab:cases}, which can explain the self-similar behavior.


\begin{figure*}[!]
\centering
\begin{subfigure}
{\includegraphics[trim = 5 0 5 0, clip,width=0.99\columnwidth]{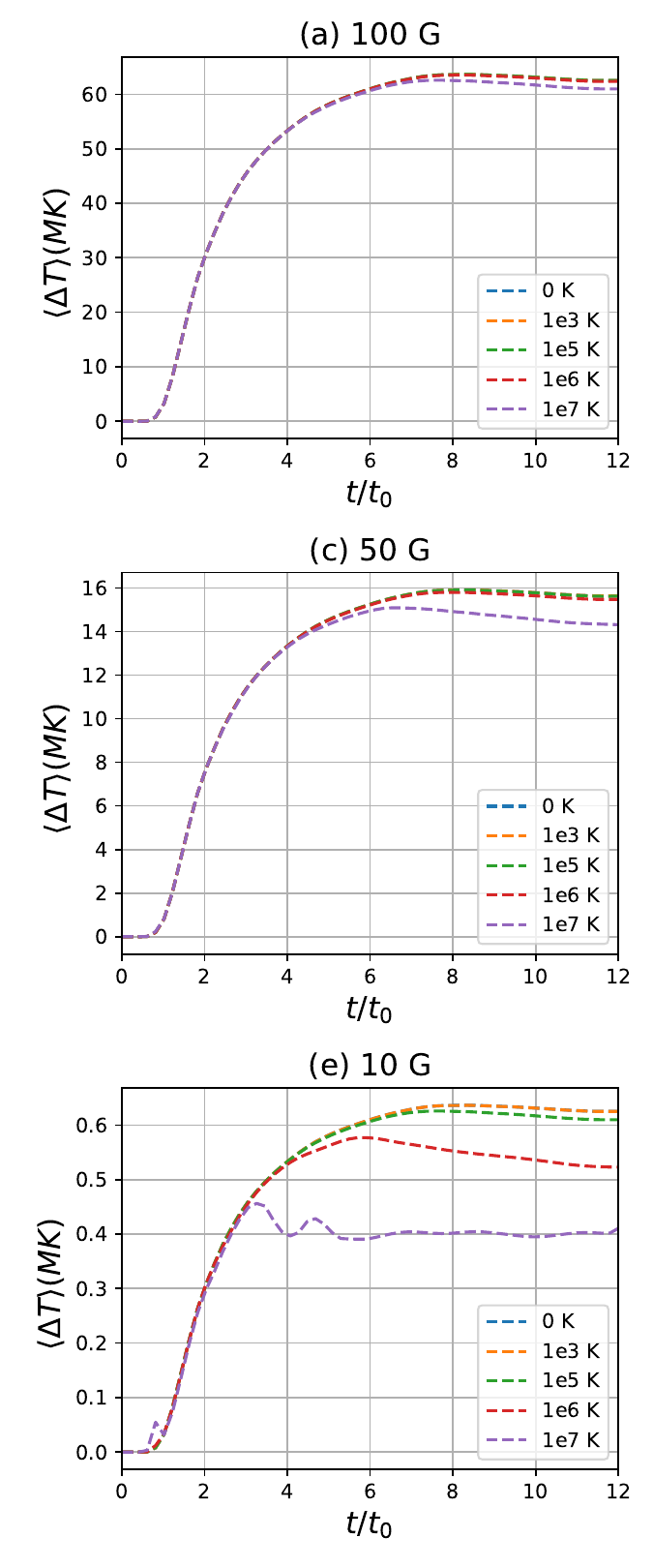}}
\end{subfigure}
\begin{subfigure}
{\includegraphics[trim = 5 0 5 0, clip,width=0.99\columnwidth]{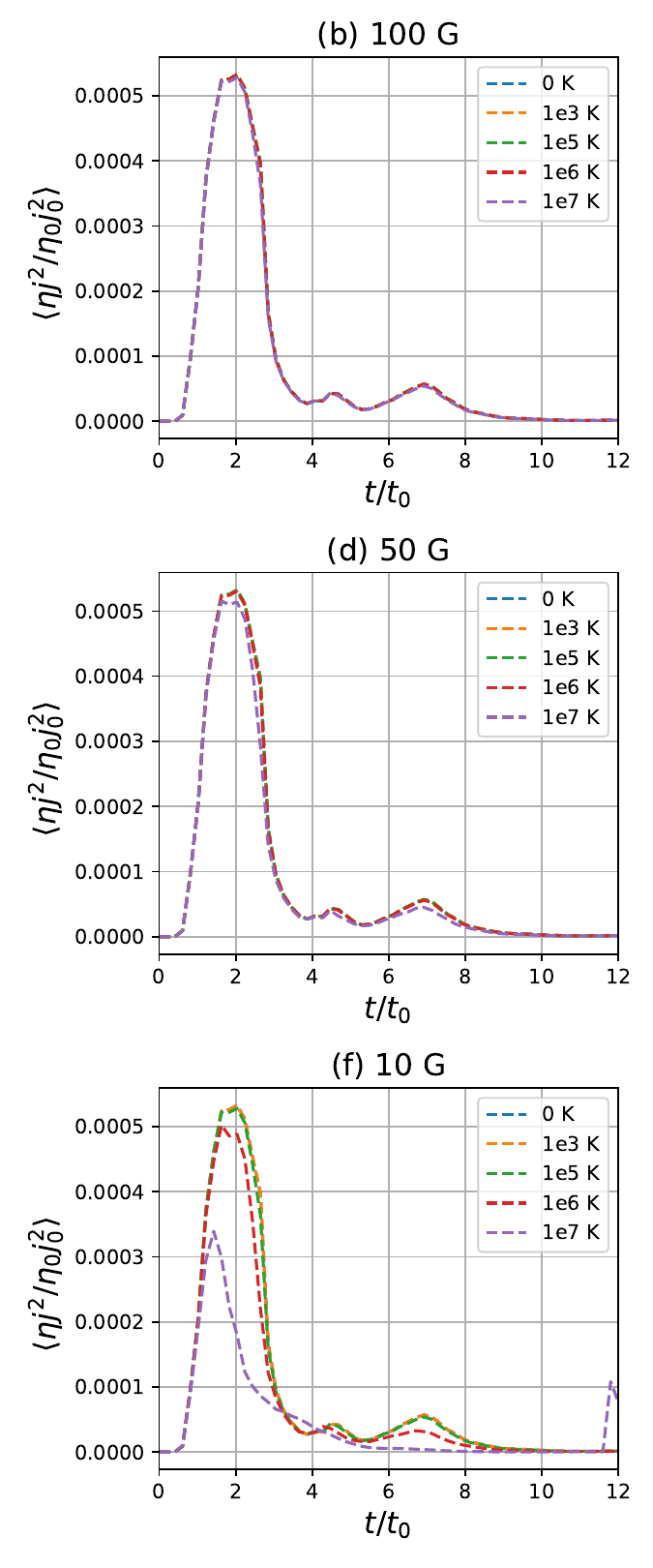}}
\end{subfigure}
\caption{Left column: Integrated temperature perturbation $\langle \Delta T \rangle$ as function of normalized time. Right column: Integrated non-dimensionalized Ohmic heating as function of non-dimensionalized time. Panels represent simulations with $B_0=$ 100 G (top), 50 G (middle) and 10 G (bottom), respectively in Table \ref{tab:cases}. }
\label{fig:heating-T}
\end{figure*}

In addition, in Figure \ref{fig:heating-loglog}b, we analyzed the maximum integrated-temperature perturbation non-dimensionalized by $T^*$ as a function of the energy ratio for multiple initial background temperatures. We observe self-similar behavior for background temperatures up to 0.1 MK. The simulation cases of 1 MK and 10 MK show a decrease in heating when $E_{i0}/E_{B0}>$  0.01.

\subsection{Sensitivity to  initial background temperature profile} \label{sec:TEMP-influence}

\subsubsection{Integrated temperature perturbation $\langle \Delta T \rangle$} \label{sec:3.3.1}


Figure \ref{fig:heating-T} displays the results of the integrated temperature perturbation $\langle \Delta T \rangle$ and integrated Ohmic heating $\langle \eta j^2 \rangle$ non-dimensionalized by $\eta_0 j^2_0$ for simulation sets B, C and D.  Firstly, analyzing the integrated temperature perturbation Figure \ref{fig:heating-T}a, significant heating is found in simulation set B, which is attributable to the high equilibrium magnetic field strength of 100 G. The curves exhibit a self-similar pattern during the initial transient, heating the plasma to a localized maximum of 65 MK in simulation set B.

Of particular interest, the heating in the simulation set at $ B_0=$100 G does not seem to be influenced by the initial temperature. This observation is significant since the initial temperature from dataset B1 and B5 differs by seven orders of magnitude (0 to 10 MK).

\begin{figure*}[!]
	\centering
 \begin{subfigure}
 {\includegraphics[trim = 10 0 10 0, clip, width=0.99\columnwidth]{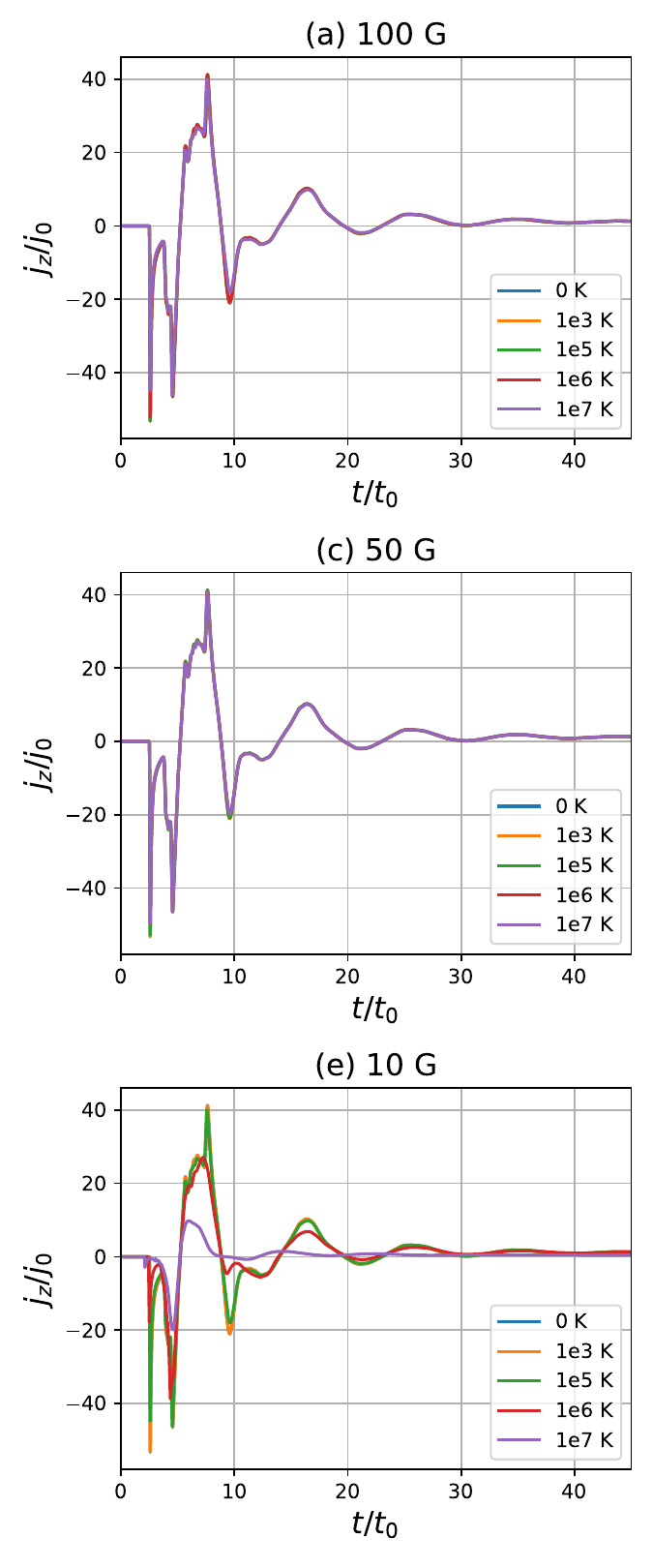}}
 \end{subfigure}
	 \begin{subfigure}
 {\includegraphics[trim = 10 0 10 0, clip, width=0.99\columnwidth]{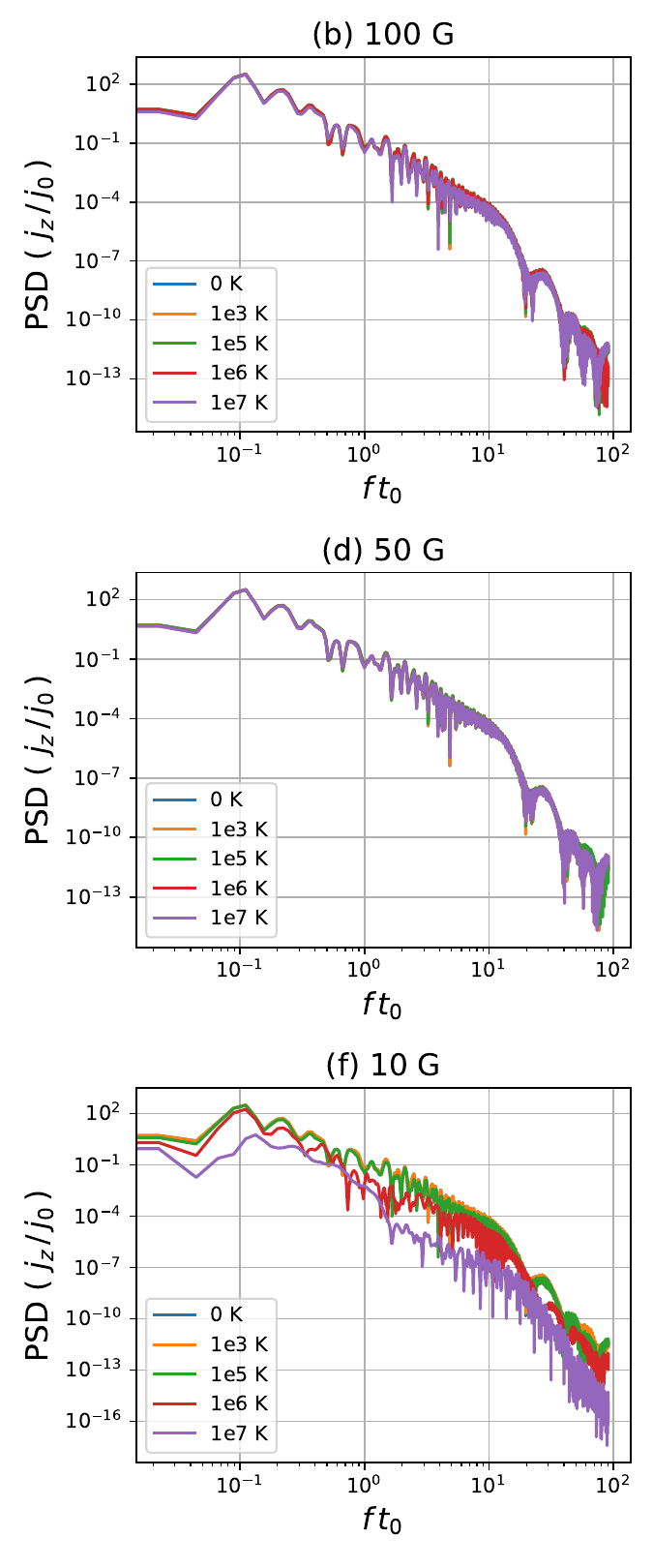}}
 \end{subfigure}
	\caption{Left column displays variations of the current density at the null point, $j_z(0,0,t)/j_0$, displays for different equilibrium magnetic field strengths and initial background temperature configurations. Right column shows the power spectral density (PSD) of $j_z(0,0,t)/j_0$ of the same simulations.}
	\label{fig:jz}
\end{figure*}

\begin{table*}[]
\centering
\caption{Periods and amplitudes extracted from $j_z(0,0,t)/j_0$ power spectral density (PSD) for the dominant period ($P$).  $\Delta P$ and  $\Delta A$ represent the period decrease and amplitude decrease from a cold case to a hot case for the same magnetic field strength.}
\label{tab:psd}
\begin{tabular}{lrrrrrrr}\hline\hline
Case & $B_0$ (G)& $T_0$ (MK)   & Period (P) & Amplitude (A) & $\Delta P$ (\%)     & $\Delta A$  (\%)    & $E_{i0}/E_{B0}$ (\%)   \\ \hline 
B1/A5 & 100 & 10    & 9.001 & 17.39 & 0.00\%  & 2.69\%  & 1.04\%   \\
B2 & 100 & 1     & 9.001 & 17.83 & 0.00\%  & 0.26\%  & 0.10\%   \\
B3 & 100 & 0.1   & 9.001 & 17.87 & 0.00\%  & 0.03\%  & 0.01\%   \\
B4 & 100 & 0.001 & 9.001 & 17.87 & 0.00\%  & 0.00\%  & 0.0001\% \\
B5 & 100 & 0     & 9.001 & 17.87 & 0.00\%  & 0.00\%  & 0.00\%   \\   \hline
C1/A4 & 50  & 10    & 9.001 & 17.69 & 0.00\%  & 1.05\%  & 4.16\%   \\
C2 & 50  & 1     & 9.001 & 17.69 & 0.00\%  & 1.05\%  & 0.42\%   \\
C3 & 50  & 0.1   & 9.001 & 17.85 & 0.00\%  & 0.10\%  & 0.04\%   \\
C4 & 50  & 0.001 & 9.001 & 17.87 & 0.00\%  & 0.00\%  & 0.0004\% \\
C5 & 50  & 0     & 9.001 & 17.87 & 0.00\%  & 0.00\%  & 0.00\%   \\   \hline
D1/A2 & 10  & 10    & 7.501 & 2.39  & 16.67\% & 86.63\% & 104.05\% \\
D2 & 10  & 1     & 9.001 & 13.36 & 0.00\%  & 25.25\% & 10.40\%  \\
D3 & 10  & 0.1   & 9.001 & 17.39 & 0.00\%  & 2.69\%  & 1.04\%   \\
D4 & 10  & 0.001 & 9.001 & 17.87 & 0.00\%  & 0.03\%  & 0.01\%   \\
D5 & 10  & 0     & 9.001 & 17.87 & 0.00\%  & 0.00\%  & 0.00\%  \\  \hline
\end{tabular}
\end{table*}

\begin{figure*}[]
	\centering
	\includegraphics[trim = 5 0 10 0, clip,width=0.98\textwidth]{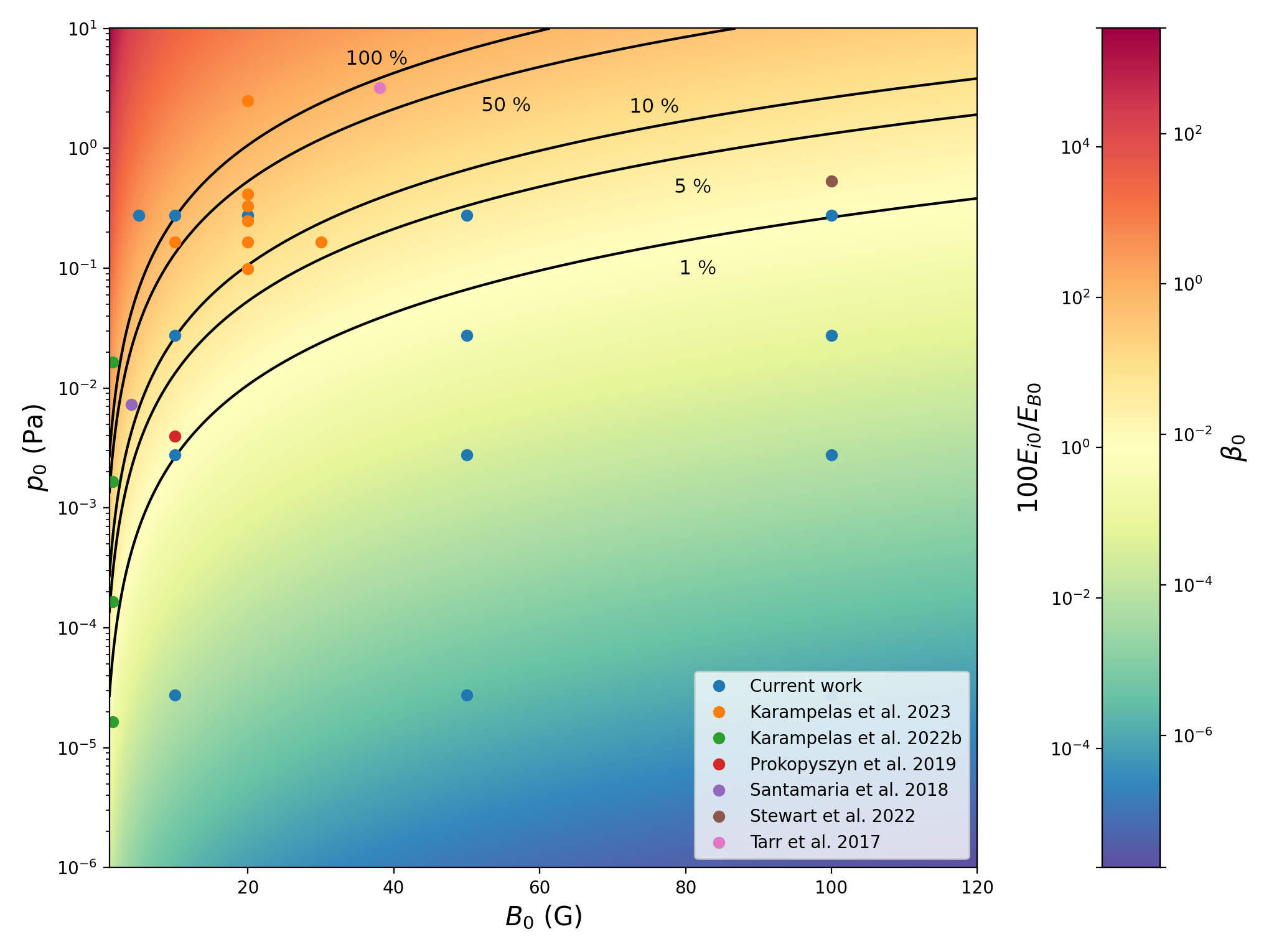} 
\caption{Contour of the ratio of internal energy non-dimensionalized by magnetic energy for the initial condition (1 Mm away from the null point). Color bar shows the energy ratio as function of plasma $\beta_0$ and $100E_{i0}/E_{B0}$. }
	\label{fig:energy-map}
\end{figure*}

In the Figure \ref{fig:heating-T}c, simulation set C (50 G) exhibits maximum heating nearly four times smaller than the results found in simulation set B  (100 G). This difference highlights the significant importance of choice of equilibrium magnetic field strength on plasma heating. Additionally, a self-similar solution is observed in temperatures between $T_0 = $ 0 to 1 MK, whereas the simulation at $T_0 =$ 10 MK achieves maximum heating faster and with a 1.5 MK reduction compared to simulations at lower temperatures.

The Figure \ref{fig:heating-T}e shows results for simulation set D (10 G). Here, the initial background temperature significantly impacts simulations, and the heating is only self-similar for simulations with temperatures ranging from 0 to 0.1 MK.
The simulation at 10 MK display a spike in $t \approx t_0$ generated by the initial shock wave; similar behavior can also be observed in $B_0=$ 5 G in Figure \ref{fig:heating-B}. 

Ohmic heating (Figures \ref{fig:heating-T}b, \ref{fig:heating-T}d and \ref{fig:heating-T}f) increases quickly, reaching its peak by $t =$ 2$t_0$, then decreasing significantly by $t =$ 4$t_0$. This increase corresponds with the sharp rise observed in the integrated temperature disturbance (Figure \ref{fig:heating-T} left column), indicating that magnetic energy is being converted to internal energy. The initial background temperature does not affect the Ohmic heating patterns observed at magnetic field strengths of 100 G and 50 G. However, simulations at 10 G reveal a decrease in the magnitude of Ohmic heating for temperatures of 1 and 10 MK, which aligns with the temperature disturbance plots in Figures \ref{fig:heating-T}a, \ref{fig:heating-T}c and \ref{fig:heating-T}e. The peak at  $t =$ 2$t_0$ suggests that a substantial portion of Ohmic heating comes from the initial jets shown in Figure  \ref{fig:flow-evo}b.

\subsubsection{Evolution of $j_z(0,0,t)$}\label{sec:oscillatory-reconnection-signal}

The evolution of the current density at the null point, $j_z(0,0,t)$, is a key aspect of oscillatory reconnection. In our simulations the null point is stationary and located at the origin $(x,y)=(0,0)$. Figure \ref{fig:jz} left column displays the $j_z(0,0,t)$ evolution (normalized against $j_0$) for datasets B, C and D in Table \ref{tab:cases} and the right column displays the power spectral density (PSD) of $j_z(0,0,t)$ non-dimensionalized. 

The observed oscillations in $j_z(0,0,t)$ are characteristic of signals associated with oscillatory reconnection, indicating periodic changes in the configuration of magnetic field lines over time \citep{McLaughlin2009}. In the Figures \ref{fig:jz}a and \ref{fig:jz}c, we see that variations in initial background temperature profile  do not significantly affect the amplitude and period of the $j_z(0,0,t)$ signal for equilibrium magnetic field strengths of $B_0 =$ 100 G (cases B) and 50 G (cases C), and these cases exhibit a self-similar solution in non-dimensional units. However, for simulations with equilibrium magnetic field strength $B_0 =$ 10 G (cases D), we observe variations in amplitude and period for temperatures of 1 MK (case D2) and 10 MK (case D1), while temperatures below 0.1 MK demonstrate a self-similar solution (and are thus comparable to $B_0 =$ 100 G and 50 G).

In Figures \ref{fig:jz}a, \ref{fig:jz}c and \ref{fig:jz}e, we can also observe that $j_z(0,0,t)/j_0$ oscillates around a non-zero value. This is due to asymmetric heating in the plasma around the neutral point, which itself is due to the sequence of reconnection jets: at the end of the simulation, the plasma to the left and right of the neutral point is fractionally hotter than the plasma above and below it, as a result of the iteration of hot reconnection jets that formed and heated the plasma (each subsequent current sheet and corresponding heating event is shorter/weaker than the last. Consequently, the plasma pressure is very slightly higher on the left and right of the null point at the very end of the simulation, making it easier for the system to form a very slight vertical current sheet). This leads to $j_z(0,0,t)$ tending towards a small positive value. This phenomenon was originally reported in 
\citet{McLaughlin2009}.

The power spectral densities (PSD) exhibit uniform behavior for all cases at magnetic field strengths of 100 G (cases B) and 50 G (cases C).  We observe a constant oscillation period for the dominant period of 9.001$t_0$ for all cases except for case D1 ($B_0$ = 10 G and $T_0$ = 1 MK). Case D1 exhibits a dominant period of 7.508$t_0$, representing a 16.7\% decrease from the other cases. We also calculated the amplitude difference for the dominant period  between a cold case and a hot case with the same magnetic field strength. We only observed a difference larger than 5\% when $E_{i0}/E_{B0}>10$\%, which represents cases D1 and D2. These results are presented in Table \ref{tab:psd}.

%


\subsection{Energy map}\label{sec:energy-map}

An explanation for the apparent independence on initial background temperature for some cases may be found in Table \ref{tab:cases}, where we observe that simulation set B operates in a low $\beta_0$ regime. The internal-to-magnetic energy ratio in simulation set B is below 0.01, indicating that magnetic energy is approximately 100 times larger than internal energy. This dominance of magnetic energy leads to self-similar solutions between a cold plasma and a plasma at 10 MK.

Let us now further explore the dependence of the internal-to-magnetic \lq{energy ratio}\rq{} on the equilibrium magnetic field strength  and the initial background temperature profile. The internal energy per unit volume, $E_i$, for an ideal fully-ionized hydrogen plasma is given by:
\begin{equation}
E_i = \rho \epsilon = \frac{2\rho k_bT}{m_p(\gamma-1)} = \frac{p}{\gamma-1} ,
\label{eq:internal}
\end{equation}
and the magnetic energy per unit volume, $E_B$, is given by:
\begin{equation}
E_B = \frac{ |{\mathbf{B}}|^2 }{2 \mu_0 } ,
\end{equation}
and thus the ratio between energies can be written as:
\begin{equation}
	\frac{E_i}{E_B} = \frac{4\mu_0 \rho k_bT}{m_p(\gamma-1) |\mathbf{B}|^2}  =  \frac{1}{\gamma-1}\frac{p}{p_{\rm{magnetic}}}  =  \frac{\beta}{\gamma-1} ,
\end{equation}  
where $p_{\rm{magnetic}}$ is the magnetic pressure.


Alternatively, it can be formulated for the initial condition at $r=L_0=$ 1 Mm of the null point:
\begin{equation}
\frac{E_{i0}}{E_{B0}} = \frac{4\mu_0 \rho_0 k_b T_0}{m_p(\gamma-1) B_0^2}  = \frac{2\mu_0}{\gamma-1}\frac{p_0}{B_0^2}  =  \frac{\beta_0}{\gamma-1}. \label{eqn:energy-ratio}
\end{equation}
 Notice  that $E_{i0}/E_{B0}$ is inversely proportional to the equilibrium magnetic field strength (at $r=$ 1 Mm), whereas it is directly proportional to density and temperature. These three variables can provide insights into whether the system is more dependent on magnetic energy or not. Notice also that in Equation (\ref{eqn:energy-ratio}) $E_{i0}/E_{B0}$ is directly proportional to $\beta_0$.



Figure \ref{fig:energy-map} denotes the internal-to-magnetic energy ratio (Equation \ref{eqn:energy-ratio}, presented as a percentage) as a function of equilibrium magnetic field strength and initial background pressure at a distance of 1 Mm from the null point. Alternatively, since the internal-to-magnetic energy ratio is directly proportional to $\beta_0$, Figure \ref{fig:energy-map} also denotes contours of $\beta_0$.  Thus, we refer to Figure \ref{fig:energy-map} as an energy map of the parameter space.

The energy map is divided into several regions delineated by isolines of $E_{i0}/E_{B0}$ in percentage at 1\%, 5\%, 10\%, 50\%, and 100\%. The contour lines provide insights into the relative dominance of internal energy compared to magnetic energy. Regions below the 10\% line suggest a dominance of magnetic energy, indicating that the system's behavior may resemble that of cold plasma, with self-similar solutions for plasma heating profiles. Solutions above the 50\% line indicate a significant contribution from internal energy, leading to a decrease in the maximum heating. Beyond the 100\% threshold, hydrodynamic effects become dominant, resulting in further reduction in heating.

In the solar corona, we can estimate the equilibrium pressure of a plasma at 1 MK and a density of $\rho_0 = 1.67\times 10^{-12}$ kg/m³ to be around 0.0276 Pa and 0.276 Pa during a 10 MK flare. Regions with pressure below this threshold are typical of the solar corona.

The solution in the lower-right-corner of the map indicate regions of low $\beta_0$ and are thus magnetically-dominated. 
The upper-left-corner indicates regions of high $\beta_0$ and so simulations here are more hydrodynamically dominated. Note that increases in thermal pressure on the $y-$axis can be obtained via increasing the initial background temperature profile and/or simulations that have higher initial background densities, according to Equation (\ref{eq:internal}). 


The energy map allows us to contextualize previous studies by placing them at specific points within the 2D parameter space:

\begin{itemize}

\item{
\cite{Karampelas2022a} investigated the influence of temperature and heat conduction on plasma conditions. Their simulations focused on a single magnetic field with varying background temperatures. They observed self-similar behavior in plasma temperature profiles at low temperatures, which is consistent with our findings.
}

\item{
\cite{Karampelas2023} conducted a parametric study covering a subset of both axes of the energy map. Their research focused on modeling $j_z$ oscillations at null points and developing empirical formulas based on their simulations. Their work primarily lies in a region above the 10\% line, which means that based on the energy map we can infer that temperature profiles may not exhibit self-similar behavior in their studied conditions.
}


\item{
The simulation from \cite{Prokopyszyn2019} is expected to display the exact solution of a cold plasma since it is placed below the $E_{i0}/E_{B0}=$ 10\% line; the same is expected from the \cite{Stewart2022} simulation. 
On the other hand, it is expected that \cite{Prokopyszyn2019} and \cite{Stewart2022} display a distinct behavior in terms of heating, since $\langle \Delta T \rangle_{\text{max}} \sim B_0^2$ as discussed in Section \ref{sec:b-influence}.
}

\item{
The simulation case from \cite{Santamaria2018} is placed above the 10\%, where the magnetic field strength is weak, meaning that magnetic and internal energies are comparable in magnitude. Their heating profile, $j_z$ oscillation period and amplitude are expected to be more sensitive to initial temperature and pressure variations. In their study, they considered an arcade configuration in a stratified atmosphere. The reference values were extracted from their plots and interpolated at the height of the null point at a 1 Mm distance for placement on the energy map.
}

\item{
\cite{Tarr2017} considered a  stratified atmospheric condition and modeled the effect of heavier ions where $\overline{m}=1.25m_p$. Their energy ratio at the height of the null point is $E_{i0}/E_{B0}=$83\%, which means that they are in a regime where plasma heating is smaller than in a cold plasma, and also it is expected to present a smaller maximum amplitude for $j_z$ signal at the null point.
}

\item{
Previous studies considering a cold plasma, $T=$ 0 K, such as \cite{McLaughlin2009} and \cite{Talbot2024} do not appear as dots on the energy map since the initial condition considers a zero pressure. Instead, they would appear towards the bottom of the $y-$axis, clearly within the low $\beta$/magnetically-dominated regime. \cite{Thurgood2017} did not provide the non-dimensionalization scales to place it into the energy map precisely. However, their energy ratio in their simulations was 1.25\%, which would place them between the 1\% and 5\% lines, which means that their simulation is in a magnetically-dominated regime.
}

\end{itemize}

Details about the initial configuration of these previous studies in the literature can be found in Section \ref{sec:intro}. As discussed, the energy map can be used to estimate the heating and $j_z$ sensitivity of a 2D X-point to different magnetic field strengths and atmospheric conditions.  





\section{Conclusions} \label{sec:conclusions}

We conducted 2D MHD simulations of a magnetic field with an X-point configuration perturbed by an initial condition in velocity for a fully-ionized, resistive plasma. Through a parametric study involving adjustments in equilibrium magnetic field strength and initial background plasma temperature, we investigated their influence on plasma heating around the null point and the oscillatory reconnection signal $j_z(0,0,t)$. 


Firstly, we performed a parameter study for different values of equilibrium magnetic field strength, $B_0$. Our equilibrium magnetic field (Equation (\ref{equation=X-point})) is scale-free and so our choice of $B_0$ is  the value of the magnetic field strength at $r=L_0$ (and then $\mathbf{B}$ varies throughout the domain). We found that the choice of $B_0$ has a significant effect on the evolution, with the maximum temperature generated by the initial reconnection jets exhibiting a quadratic dependence $\langle \Delta T \rangle_{\text{max}} \sim B_0^2$. For example, our simulation for $B_0$ = 50 G and $T=$ 10 MK (case C1) generated reconnection jets with maximum temperatures of around 15 MK. We also obtained an expression to quantify the heating and the average temperature as $\langle T \rangle_{\text{max}} =  a B_0^{b} + T(t=0)$, with Table \ref{tab:fitting} detailing the amplitude $a$ and index $b$ for all our simulations. 

This behavior aligns with expectations, as the magnetic energy increases quadratically with the magnetic field strength, resulting in a larger reservoir of magnetic energy available for conversion into internal energy, creating a self-similar behavior of Figure \ref{fig:heating-loglog}b.


We also analyzed the maximum integrated-temperature perturbation (non-dimensionalized by $T^*$) as a function of the internal-to-magnetic energy ratio for multiple initial background temperatures. We found self-similar behavior for all cases where $E_{i0}/E_{B0}<$  0.01, including self-similar behavior for all background temperatures up to 0.1 MK, and a decrease in heating for simulation cases of
10 MK (case D1) and 1 MK (case D2).



Secondly, we performed a parameter study for different values of initial background temperature $T_0$. We found that the  integrated temperature perturbation $\langle  T(x,y,t) - T_0 \rangle$ displayed different behavior depending upon the value of $B_0$. For $ B_0=$ 100 G (cases B) and $ B_0=$ 50 G (cases C),  the integrated temperature perturbation did not seem to be influenced by the initial background temperature; all the curves exhibit a self-similar pattern, over seven orders of magnitude (0 to 10 MK) for cases B and six orders of magnitude (0 to 1 MK)  for cases C: there was only a slight departure from self-similarity for case C1 ($T_0$ = 10 MK).

For $B_0$ = 10 G simulations (cases D), we found that the evolution of the integrated temperature perturbation is only self-similar for simulations with temperatures ranging from 0 to 0.1 MK (cases D3 to D5), whereas for 10 MK (case D1) and 1 MK (case D2) the initial background temperature does significantly impact the evolution (a departure from self-similarity).

Although the heating effect is more pronounced in simulations at lower temperature, where $\beta_0$ is smaller, the actual temperature is larger in simulations when the initial background plasma is hotter. This is expected since final temperature is still dependent on the initial temperature in the system, as seen in Equation (\ref{eq:dt}).

We also investigated the amount of Ohmic heating in the system, finding that the initial background temperature does not affect the integrated Ohmic heating patterns observed at magnetic field strengths of 100 G (Cases B) and 50 G (Cases C). However, simulations at 10 G reveal a decrease in the magnitude of Ohmic heating for temperatures of 10 MK (case D1) and 1 MK (case D2), in agreement with the temperature disturbance plots.


Thirdly, we also investigated the  evolution of the current density at the null point, $j_z(0,0,t)$ for different values of initial background temperature $T(t=0)$. We found that  variations in initial background temperature profile  do not significantly affect the amplitude and period of the $j_z(0,0,t)$ signal for equilibrium magnetic field strengths of $B_0 =$ 100 G (cases B) nor 50 G (cases C), and that all these cases exhibit a self-similar solution in non-dimensional units. However, for simulations with $B_0 =$ 10 G (cases D) we observe variations in amplitude and period for temperatures of 10 MK (case D1) and 1 MK (case D2), while temperatures below 0.1 MK demonstrate a self-similar solution (and are thus comparable to $j_z(0,0,t)$ for $B_0 =$ 100 G and 50 G), i.e. a similar result as that for the integrated temperature perturbation sensitivity.

The power spectral densities (PSD) exhibit uniform behavior for all cases at magnetic field strengths of 100 G (cases B) and 50 G (cases C) and for 10G Cases D2-D5, where we observed a constant oscillation period for the dominant period of 9.001$t_0$ for all cases except for case D1 ($B_0$ = 10 G and $T_0$ = 1 MK). Case D1 exhibits a dominant period of 7.508$t_0$, representing a 16.7\% decrease from the other cases.

Additionally, we calculated the amplitude difference for the dominant period between a cold case and a hot case with the same magnetic field strength. We only observed a difference larger than 5\% when $E_{i0}/E_{B0}>10$\%, which represents cases D1 and D2.


Fourthly, the results for the parameter studies across $B_0$ and $T(t=0)$ let us visualize the 2D parameter space as an energy map (Figure \ref{fig:energy-map}). This energy map presents the ratio of internal-to-magnetic energy $E_{i0}/E_{B0}$ or, equivalently, plasma $\beta_0$, since ${E_{i0}} / {E_{B0}}=  {\beta_0} / {(\gamma-1)}$.

The energy map is divided into several regions delineated by isolines of $E_{i0}/E_{B0}$, where contour lines provide insights into the relative dominance of internal energy compared to magnetic energy or, equivalently, in terms of $\beta_0$:
\begin{itemize}
\item
{Regions below the 10\% line suggest a dominance of magnetic energy, with self-similar solutions for plasma heating profiles and for $j_z(0,0,t)$. Equivalently, the lower-right-corner of the map indicates regions of low $\beta_0$ and are thus magnetically-dominated; this is where cold plasma simulations exist.
}        
\item{
Solutions above the 50\% line indicate a considerable contribution from internal energy, leading to a decrease in the maximum heating. 
}
\item{
Beyond the 100\% threshold, hydrodynamic effects become dominant, resulting in further reduction in heating. Equivalently, the upper-left-corner indicates regions of high $\beta_0$ and so simulations here are more hydrodynamically dominated. There is a significant departure from the self-similar solutions of the magnetically dominated regimes.
}   
\end{itemize}

The energy map also allows us to contextualize previous studies by placing them at specific points within the 2D parameter space (see Figure \ref{fig:energy-map}).  All these previous studies reported an oscillating signal in $j_z(0,0,t)$ but also reported a variety of periods, amplitudes and behaviors for such a signal, and the energy map now brings these different studies together into a  single unified understanding (i.e. there exists a 2D parameter space, dependent upon choices of $B_0$ and $T(t=0)$, that can affect resultant behavior and departure from self-similar solutions in extreme high $\beta_0$ situations). Thus, this energy map serves as a valuable tool for interpreting plasma behavior within a broader parameter space.

It is important to mention that these findings rely on the hypothesis that the plasma is fully ionized, and thermal conduction and radiation effects were not considered in this study (\citealt{Karampelas2022a} investigated the effect of thermal conduction of the oscillatory reconnection system, finding that its inclusion has a small effect on the resultant period).


\section*{Acknowledgments}
We would like to thank the anonymous referee for the constructive comments that improved the quality of the paper.
All authors acknowledge the UK Research and Innovation (UKRI) Science and Technology Facilities Council (STFC) for support from grant ST/X001008/1 and for IDL support.
This work used the DiRAC Data Intensive service (CSD3) at the University of Cambridge, managed by the University of Cambridge University Information Services on behalf of the STFC DiRAC HPC Facility (www.dirac.ac.uk). The DiRAC component of CSD3 at Cambridge was funded by BEIS, UKRI and STFC capital funding and STFC operations grants. DiRAC is part of the UKRI Digital Research Infrastructure. Numerical simulations were conducted with LARE2D which is available at \href{https://github.com/Warwick-Plasma/Lare2d}{https://github.com/Warwick-Plasma/Lare2d}. The data behind Figure \ref{fig:flow-evo}, and a python script to reproduce it, are available in the following repository, DOI: 
\dataset[10.25398/rd.northumbria.26310742]{https://figshare.com/s/668c32b6c51747df0b07}.



\software{LARE2D \citep{Arber2001},
          NumPy \citep{numpy},  
          SciPy \citep{SciPy}, 
          matplotlib \citep{matplotlib}.
          }


\bibliography{references}{}
\bibliographystyle{aasjournal}



\end{document}